\documentclass[english]{article}
\usepackage[T1]{fontenc}
\usepackage[latin9]{inputenc}
\usepackage{array}
\usepackage{multirow}
\usepackage{amstext}
\usepackage{graphicx}
\usepackage{xcolor}
\usepackage{url}
\usepackage{tabularx}
\makeatletter
\usepackage{blindtext}
\usepackage{enumitem}

\usepackage[T1]{fontenc}
\usepackage[latin9]{inputenc}
\usepackage{array}

\usepackage{xcolor}

\usepackage{graphicx}
\usepackage[compatibility=off]{caption}
\usepackage{subcaption}

\setkeys{Gin}{width=\linewidth}

\makeatother

\usepackage{babel}

\usepackage{geometry}
\begin{document}
	\begin{flushleft}
		{\Large
			\textbf\newline{Discovery of the Twitter Bursty Botnet} 
		}
		\newline
		\\
		Juan Echeverria$^{1*}$,
		Christoph Besel, 
		Shi Zhou$^{2*}$
		
		\bigskip
		Department of Computer Science\\University College London (UCL), London,  United Kingdom
		\\
		$^{1*}$ \url{j.guzman.11@ucl.ac.uk} \\
		$^{2*}$ \url{s.zhou@ucl.ac.uk}
		
\end{flushleft}

\begin{abstract}
	
	Many Twitter users are bots. They can be used for spamming, opinion manipulation and online fraud. 
	Recently, we discovered the {\em Star Wars botnet}, consisting of more than 350,000 bots tweeting random quotations exclusively from Star Wars novels. 
	The bots were exposed because they tweeted uniformly from any location within two rectangle-shaped geographic zones covering Europe and the USA, including sea and desert areas in the zones. 
	In this paper, we report another unusual  behaviour of the Star Wars bots, that the bots were created in bursts or batches, and they only tweeted in their first few minutes since creation. 
	Inspired by this observation, we discovered an even larger Twitter botnet, the {\em Bursty botnet} with more than 500,000 bots.  
	Our preliminary study showed that the Bursty botnet was directly responsible for a large-scale online spamming attack in 2012. 
	Most  bot detection algorithms have been based on assumptions of `common' features that were supposedly shared by all bots. Our discovered botnets, however, do not show many of those features; instead, they were detected by their distinct, unusual tweeting behaviours that were unknown until now. 

\end{abstract}

\section{Introduction}
	Twitter bots are Twitter user accounts created and controlled by hackers, called botmasters, using computer programs. A Twitter botnet is a group of  bots that show the same properties and are controlled by the same botmaster.
	The Twitter company has identified and removed  millions  of bots. Researchers claim there are much more bots on Twitter \cite{morstatter_fred_asonam}, and they have introduced a number of methods to detect Twitter bots.
	
	Twitter bots can pose a series of threats to cyberspace security~\cite{morstatter_fred_asonam}. For example, they can send a large amount of  spam tweets to other users; they can create fake treading topics; they can manipulate public opinion; they can launch a so-called astroturfing attack where they orchestrate false `grass roots' campaigns to create a fake sense of agreement among Twitter users \cite{ferrara_manipulation_2015,ratkiewicz_truthy:_2011,abokhodair_dissecting_2015}; and they can contaminate the data from Twitter's streaming API \cite{morstatter_can_2016} that so many research works  have been based on; they have even been linked to election disruption \cite{bessi_social_2016}.

Recently we discovered the {\em Star Wars botnet} \cite{echeverria_star_2017} with more than 350,000 bots. 
These bots were discovered because they tweeted uniformly from any location within two rectangle-shaped geographic zones covering Europe and the USA, including sea and desert areas in the zones. 
Further inspection showed the bots only tweeted random quotations from Star Wars novels. 
This botnet was discovered and detected  in a way completely different from previous bot detection efforts.

In this paper, we report our discovery of another Twitter botnet, the {\em Bursty botnet}, which is even larger -- with more than 500,000 bots that have not been banned or removed  by Twitter the time of writing.  
The discovery of the Bursty botnet was inspired by another unusual tweeting behaviour of the Star Wars bots. 
Our preliminary study showed that the Bursty botnet was directly responsible for a large-scale online spamming attack in 2012. 
			 
	 Our work not only provides valuable ground truth data for research on Twitter bots, but also enabled us to reflect on the limitations of existing methods for detecting Twitter bots.  
	 Existing methods are mostly based on  `common' features that were supposedly shared by all Twitter bots. Our newly discovered botnets, however, do not show any of those features; instead, they were detected by their distinct, unusual tweeting behaviours that were unknown until now.

	\section{Background}
			 
			 \subsection{Twitter bots detection}
			 
			 Twitter, as a company, has been actively identifying and removing suspicious users, many of which are spammers or bots \cite{thomas_suspended_2011}. 
			 Researchers have also proposed various methods to classify or detect Twitter bots, many of which are machine learning methods.   
			 Many of these works were based on features either measured from ground truth data or assumed by researchers as common properties of all bots.  
			 Such features include user properties, including username length \cite{lee_early_2014},  user profile \cite{zafarani_10_2015}, time between tweets \cite{chu_who_2010}, Levensthein distance between the tweets of a user \cite{wang_detecting_2010}, distribution of  tweeting frequency \cite{chu_who_2010} and entropy in tweeting frequency \cite{dickerson_using_2014}; as well as tweet text properties \cite{chu_who_2010}, including   topics  \cite{dickerson_using_2014} and sentiment analysis  \cite{dickerson_using_2014}.  
			 Some researchers claimed that their bot classifiers can achieve high accuracy, as high as 99.99\% \cite{thomas_trafficking_2013}. 
			 
			 The existing efforts on Twitter bots detection, however,   suffer a major problem, that 
			 despite their ever more sophisticated algorithms, they could only discover  small numbers of (mixed types of) bots, mostly hundreds or up to a few thousands -- and yet it is well known that there are very large  botnets on Twitter  \cite{narang_satnam_green}.

			 The performance of bot detection methods is ultimately determined by our knowledge and understanding of Twitter bots. However,  there is a well acknowledged lack of ground truth data \cite{subrahmanian_darpa_2016}. Twitter's bot datasets are not available to the public. The small number of available datasets are small and contain mixed types of bots \cite{morstatter_fred_asonam}. 
			 As a result, it is not clear which assumptions on Twitter bots are true,  which features are most characteristic, or  
			 whether Twitter bots should have common features at all.

			\begin{figure}[tbh]
		\centering
		\includegraphics [width=15cm] {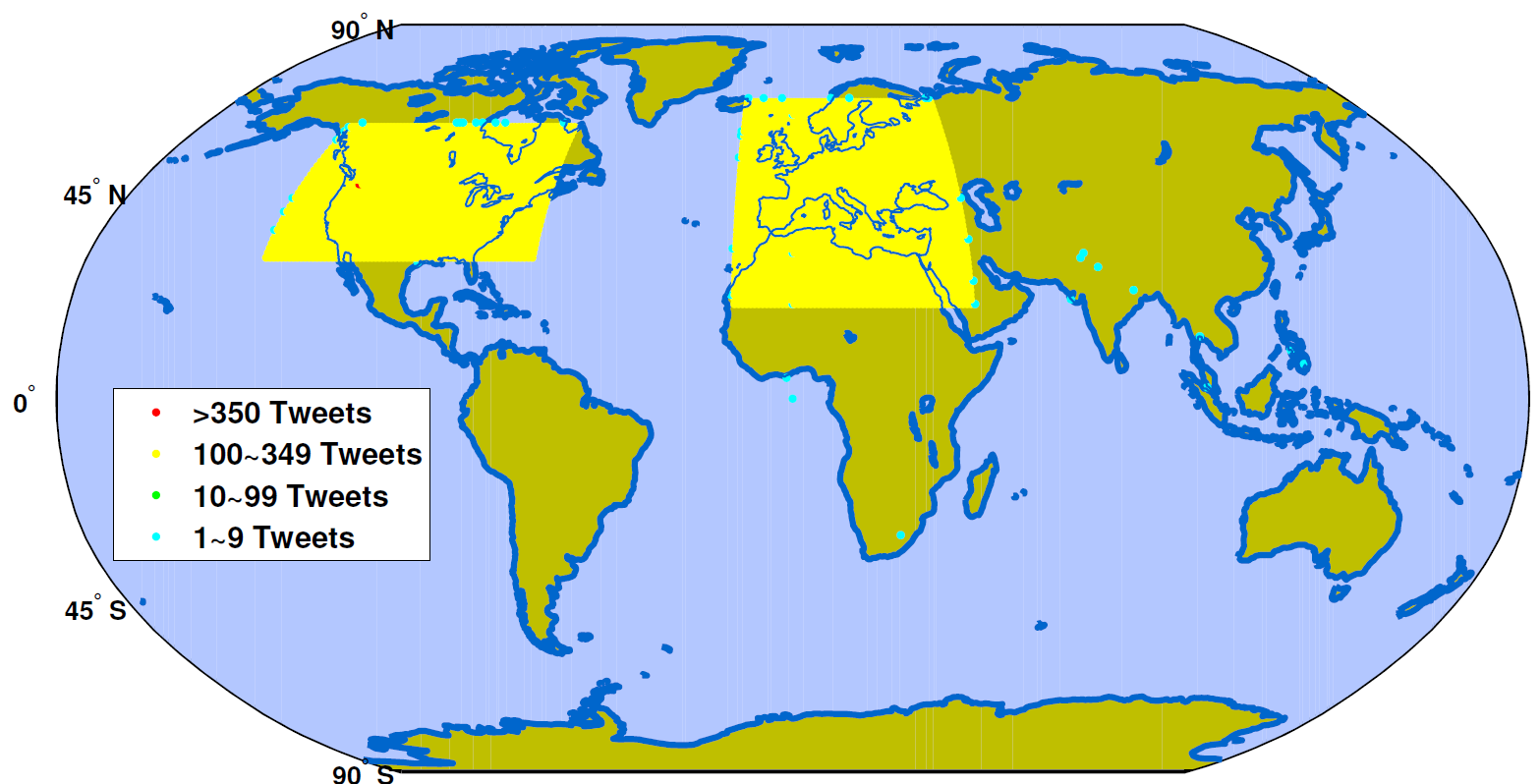}	
		\caption{Distribution of tweet locations of the Star Wars bots. The bots exclusively, uniformly tweet from any location within the two rectangle zones over Europe and North America, including sea and desert areas. This unusual tweeting behaviour was the first clue for discovering the Star Wars bots. }\label{starwars}
	\end{figure}
		 
\subsection{The Star Wars botnet}
	Recently we discovered a large botnet, the Star Wars botnet \cite{echeverria_star_2017}. This discovery stems  from the fact that the Star Wars bots create clear rectangular patterns when plotting their location-tagged tweets on a world map. After manual tagging of some of the bots, a Naive Bayes classifier was trained and some filters were applied. Finally, more than 350,000 Star Wars bots were identified showing the following properties. 	
	\begin{itemize}		
		
		\item  The Star Wars bots exclusively and textually tweet quotations from Star Wars novels, with the only exception of inserting the hash character (\#) or special hashtags (such as  {\#followme}) at random places.  
		
		\item The bots  were registered with Twitter   from June to July 2013, and therefore their user ID numbers  are within a narrow range between  $1.5 \times 10^9$ and $1.6\times 10^9$. 
		
		\item The tweeting source of the bots is exclusively 'Windows Phone', which  accounts for only 0.02\% of all tweets on Twitter. 
			
		\item About half of all tweets of the bots are location-tagged. Almost all of the bots have tweeted at least one location-tagged tweet. The tweet location tags are obviously fake because all of them fall in one of the two rectangle zones, uniformly (see Figure~\ref{starwars}).
		
		\item Because of the fake locations, distance between two consecutive geo-located tweets of a bot is very large, over 2000 km on average.
		
		\item The bots  have created no more than 11 tweets in their
lifetime; and they have no retweets or mentions. 
		
		\item The Star Wars bots disproportionately follow other Star Wars bots (the botnet follows itself). They have at most 10 followers and at most 31 friends.  
		
	\end{itemize}
 
	The Star Wars botnet provides a valuable dataset for studying Twitter bots. It is not only very large, but also contains a single botnet showing the same properties. It provides rich information and clues on how the botnet is designed and created \cite{echeverria_star_2017}. The Star Wars bots do not show many of the previously assumed `common' features of Twitter bots; instead they exhibit a number of unusual tweeting behaviours that have not been reported. The discovery of the botnet was accidental, but it   illustrates the limitations of existing bot detection methods.  

\begin{figure*}
	\centering
	
	\begin{subfigure}[b]{18cm}
		\hspace{-1.5cm}\includegraphics [width=18cm]{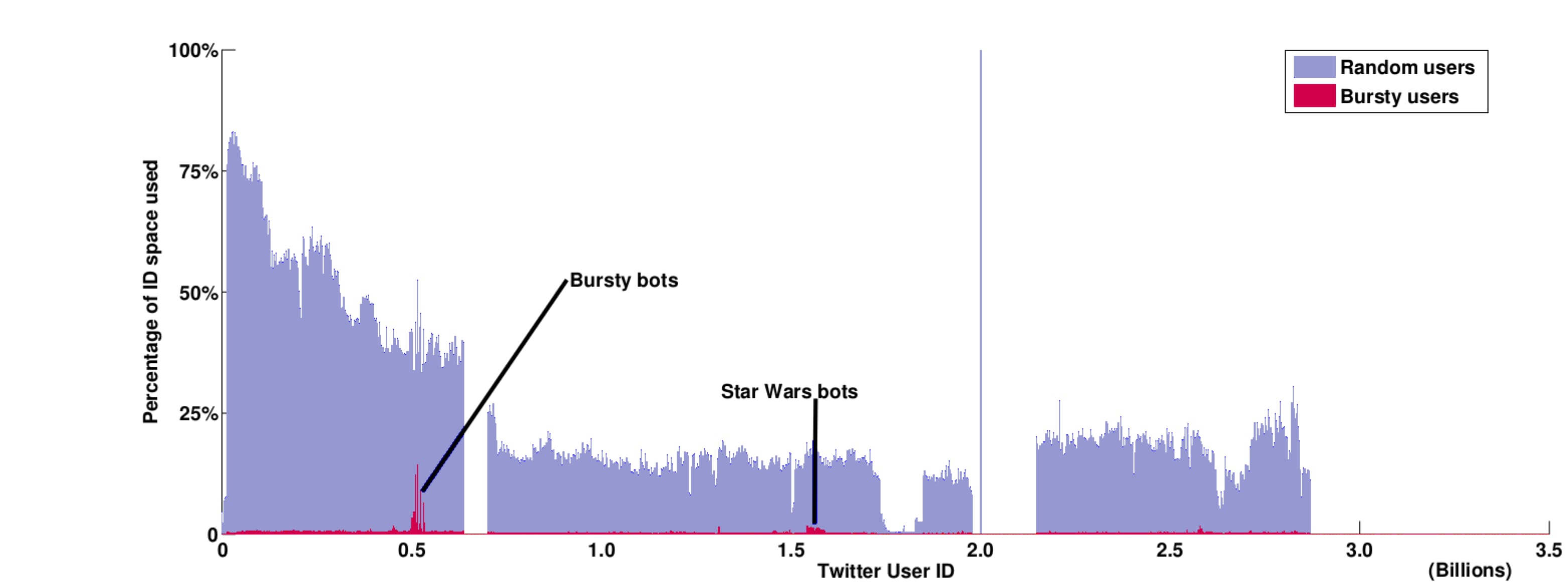}
		\caption{Percentage of Twitter user IDs allocated to all users and bursty users in each bin of 1 million IDs.}				
	\end{subfigure} 	
	
	\vspace{1cm}
	
	\begin{subfigure}[b]{18cm}
		\hspace{-1.5cm}\includegraphics[width=18cm]{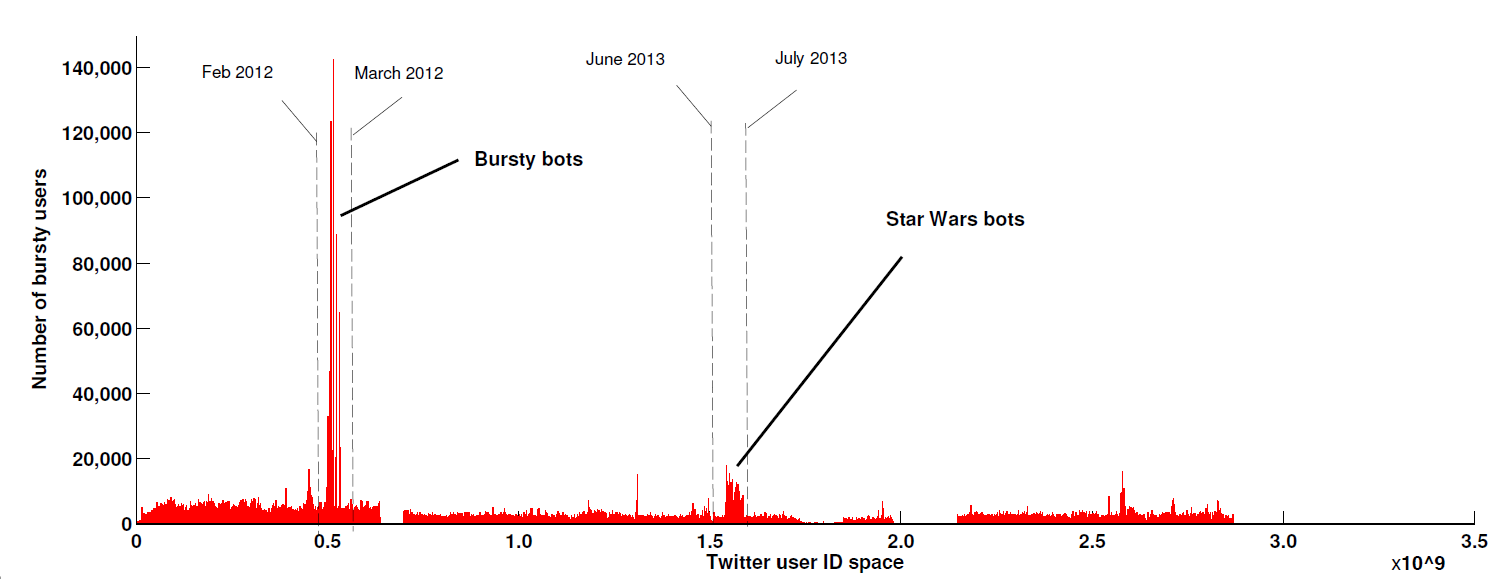}
		\caption{Number of user IDs allocated to bursty users.   }
	\end{subfigure} 
	
	\caption{Distribution of   the bursty users in the Twitter user ID space in September 2015.  A blue line in (a) near 2.0$\times10^9$ are IDs reserved by Twitter.   }
	\label{fig: Bursty Users}
\end{figure*}

\section{Discovery of the Bursty botnet}

In this section we will show how we detect another botnet, with over 500,000 users. This  was inspired by an unusual tweeting behaviour that we observe on the Star Wars bots: all the bots only tweeted in the first few minutes after their creation.  
We call it the bursty tweeting behaviour. 
We define  the  {\em bursty} users  as those who tweeted  at least 3 times in the first hour following user creation, and then never tweeted again.     
Figure \ref{fig: Bursty Users} shows the distribution of   bursty users in the Twitter user ID space.  When a Twitter user is created,  Twitter assigns the user a unique  ID number between 0 and $2^{32}$. In general, the user IDs are allocated in a sequential manner such that a smaller ID means an earlier creation date.  

We can see in Figure \ref{fig: Bursty Users}(a) the bursty users are a tiny fraction of Twitter users.  As shown in Figure \ref{fig: Bursty Users}(b), the Star Wars bots can be clearly identified as a cluster of spikes with IDs in the range between  $1.5 \times 10^9$ and $1.6\times 10^9$, which corresponds to  June and July 2013, the period when the Star Wars bots were created. This is as expected, because Star Wars bots are bursty users. 

However, in Figure \ref{fig: Bursty Users}(b), there is another much more prominent cluster of spikes. These bursty users were created in February and March 2012 within the ID range between $5.00\times10^8$ and $5.35\times10^8$. 
We manually checked these users and noticed that many of them showed clear characteristics of spamming bots because their tweets contained only a mention and/or a URL. Many of the URLs are shortened and pointed to blocked domains. An example of a bursty bot can be seen in Figure \ref{fig:  BurstyBot}.
Further study showed that these suspicious users were members of an unknown botnet, we call it  the {\em Bursty} botnet. 


\begin{figure}
	\centering
	\includegraphics[width= 10cm]{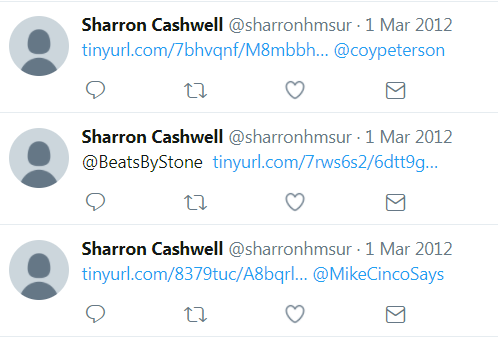}
	\caption{Screenshot of a Bursty Bot live on Twitter.  }
	\label{fig: BurstyBot}
\end{figure}

\subsection{Definition of the Bursty bots}
Here we define the Bursty bots as Twitter users with all of the following properties. 

		\begin{itemize}			
			\item They were registered in February and March 2012 with user IDs between $5\times10^8$ and $5.35\times10^8$.  
			\item They show bursty tweeting behaviour. That is, they generated at least three tweets and they only tweeted in the first hour after their creation.   
			\item They only tweeted from the source of 'Mobile Web'. 
			\item They mostly tweet (i) a URL; or/and (ii) a mention of another user.
		\end{itemize}
We  retrieved   about 21 million users in the ID range of $5\times10^8$ and $5.35\times10^8$.  We collected all of their user account information and all their tweets up to September 2016. According to the above definition, we identified more than 500,000  Bursty bots.



\subsection{Bursty tweeting and bursty creation}
\begin{figure}[h]
	\centering
	\includegraphics[width=8cm]{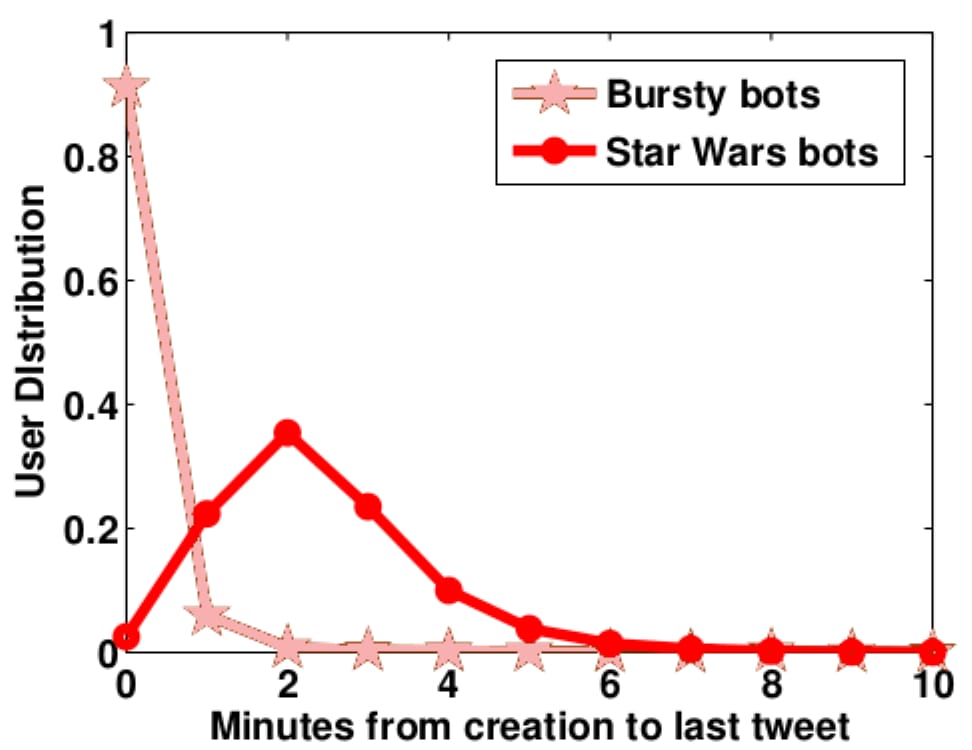}
	\caption{Distribution of the Bursty bots and Star Wars bots as a function of   time (in minutes)  from their  creation   to their last  tweet.  Most of the bots only generated tweets in their first few minutes. }
	\label{fig: UserCreationToLastTweet}
\end{figure}

The Bursty botnet shows two bursty properties: they tweeted in a bursty way and they were  created in bursts.  
Figure \ref{fig: UserCreationToLastTweet} shows most the Bursty bots and the Star Wars bots only tweeted in the first few minutes since their creation.  In particular, the Bursty bots show a strong bursty behaviour that almost all tweets were generated less than 2 minutes after account creation and then stayed silent forever. 
This is a clear sign of automatised behaviour.   This feature also gives us a sharp time frame to look for the possible cyber attacks that the bots were created for.    

Figure \ref{fig:BurstyUsersMinusBurstyBots_OnlyEnglish} shows that most of the bursty users created in February and March 2012, with user IDs from 500$\times10^6$ to 535$\times10^6$, are identified as the Bursty bots, showing all the properties defined above. 
Only a small, stable  number of bursty users are not identified as Bursty bots. These are perhaps normal users who accidentally  joined Twitter at that time, tried a few tweets and then never return again.  Such users appear with a low and almost constant rate throughout the time. 

Although we defined the bursty users as those who  tweeted only in the first hour, we now know that most of the Bursty Bots actually tweeted only in the first two minutes. The fact that almost all bursty users can be identified as the Bursty Bots means that bursty tweeting is indeed an unusual, distinct behaviour of this  botnet.

\subsection{The `disappeared' Bursty bots}

\begin{figure}
	\centering
	
	\begin{subfigure}[b]{18cm}
		\centering
		\includegraphics[width=8.5cm]{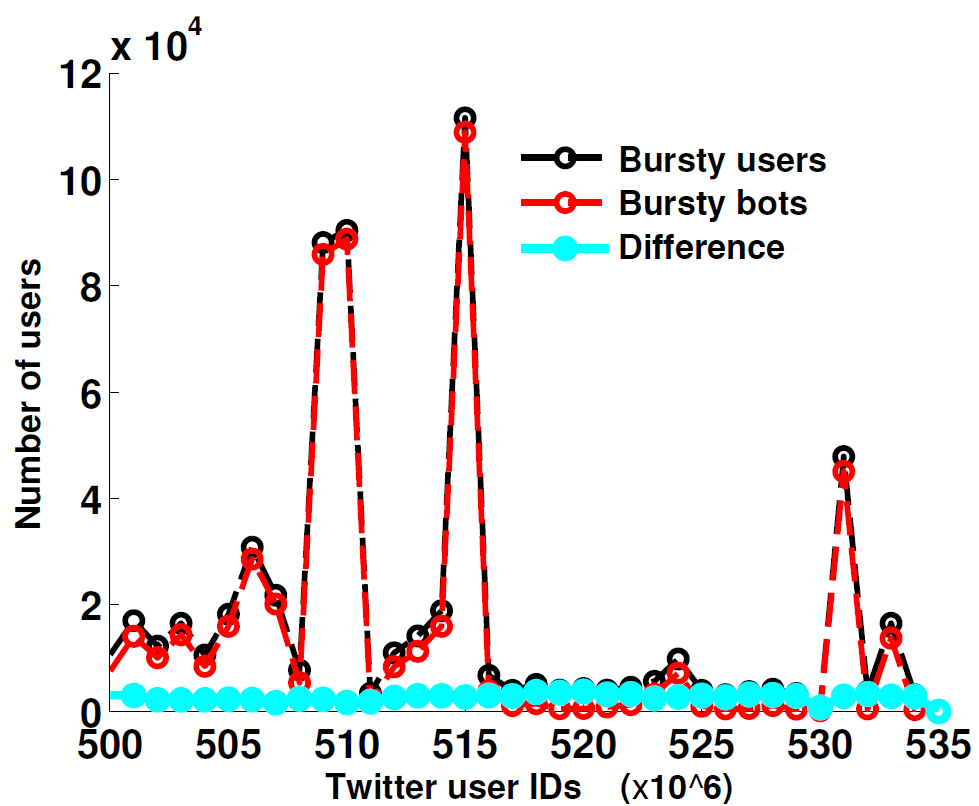}
		\caption{Bursty users and the identified Bursty bots as measured in September 2016. Most bursty users are Bursty bots. }
		\label{fig:BurstyUsersMinusBurstyBots_OnlyEnglish}			
	\end{subfigure} 	
	
	\vspace{1cm}
	
	\begin{subfigure}[b]{18cm}
		\centering
		\includegraphics[width=8.5cm]{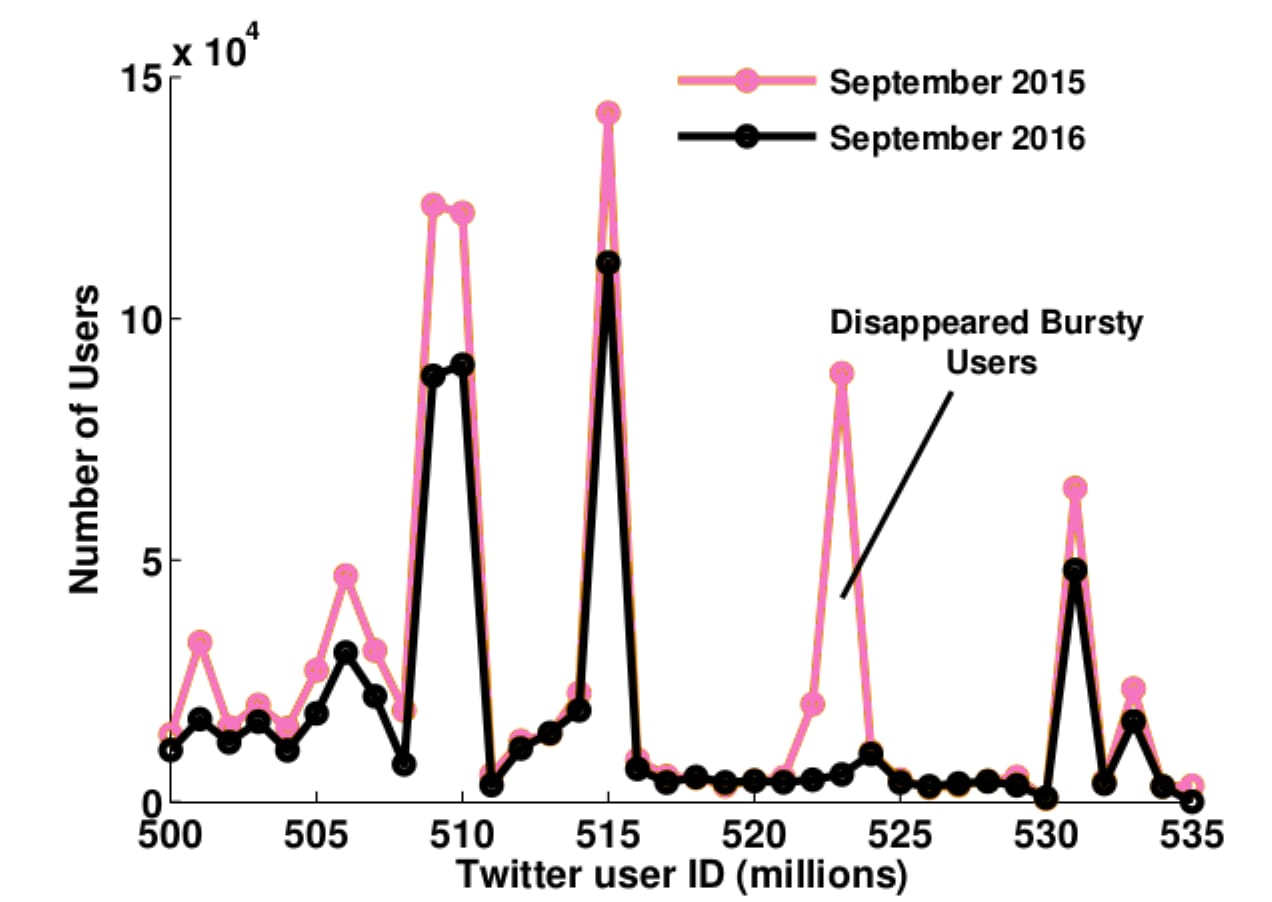}
		\caption{Bursty bots  in September 2015 and September 2016. Many bursty bots disappeared during that time.   }
		\label{fig:BurstyUsers1pcAnd100pc_OnlyEnglish}
	\end{subfigure} 
	
	\caption{Bursty users and the Identified Bursty bots in the Twitter user ID range of 500$\times10^6$ and 535$\times10^6$. }
	\label{fig:  cc}
\end{figure}

We  collected the bursty users in September 2015 and again in September 2016. 
As shown in Figure \ref{fig:BurstyUsers1pcAnd100pc_OnlyEnglish}, about 300,000 Bursty bots have disappeared during that period. 
Notably, there is a whole spike of Bursty bots missing with IDs between $520 \times10^6$ and $525 \times10^6$. It is likely that their accounts were removed by Twitter, we checked many of their accounts, which are indeed suspended.
On one hand, this supports our detection of the Bursty bots as computer-controlled, malicious users; on the other hand, it shows that Twitter has not identified the Bursty botnet as a whole, leaving the majority of the botnet still alive online.

\subsection{Unusual connectivity}

Table \ref{tab: bursty stat table} shows the Bursty bots have a number of distinct properties. Most of the bots do not have any followers (outgoing links) or friend (incoming links). 
It is notable that the definition of the Bursty Botnet did not involve user connectivity, yet the detected Bursty Bots exhibited such an unusual feature. 

This  feature is against a popular assumption in previous  studies that Twitter bots should tend to have many connections.
This feature, however, is  expected from the bursty creation of the Bursty Bots, as they were designed to be used only once immediately after they were created. 
%


\begin {table}
\centering
\begin{tabular}{|l|r|}
	\hline
	\textbf{Property}&\textbf{Value or percentage}\\\hline\hline
	
Bots with no friend &99.0\%\\\hline
Bots with no follower&99.0\%\\\hline	
Bots with tweets with URLs &98.0\%\\\hline
Bots with tweets with mentions &97.5\%\\\hline
Bots with tweets with hashtags &0.8\%\\\hline\hline

Average number of tweets &4.74\\\hline
Total number of tweets & 2.8 million\\\hline
Tweets with URLs&97.6\%\\\hline		
Tweets with mentions &64.1\%\\\hline
Tweets with hashtags&2.7\%\\\hline

\end{tabular}
\caption{Properties of the Bursty bots. }
\label{tab: bursty stat table}
\end{table}


\begin{table}[h]
	\centering
	\begin{tabular}{|r|r|}
		\hline
		\textbf{Domain}&\textbf{Count}\\\hline\hline
		tinyurl.com&1,179,369\\\hline
		google.com&562,557\\\hline
		bit.ly&328,016\\\hline
		dietagolder670.ru&54,585\\\hline
		goroskopsiris2346.ru&54,414\\\hline
		dietagoliu4758.ru&52,992\\\hline
		dietaseru858.ru&51,894\\\hline
		
	\end{tabular}
	\caption{Domains most tweeted by the Bursty bots}
	\label{tab: domains}
\end{table}

\subsection{The Bursty botnet spamming attack}

It is notable that almost all of the tweets generated by the bots contain a URL; and about 2/3  have a mention. This means that almost all of the tweets that have a mention also have a URL.  This indicates that the bots were likely created for spamming attacks, the mentions were used to maximise the reach of the tweet, both by attracting the user being mentioned and his followers to click on the URL.

To find out more details, we  examined all the URLs tweeted by the Bursty bots.  
Of the 2.8 million tweets that the Bursty botnet has created, almost all (over 99.9\%) of the URLs were unique, which means most of the URLs were only tweeted once by a single bot. 
%
As shown in Table \ref{tab: domains}, when divided by domain,  the most tweeted domain is the URL shortening and redirect service, \url{tinyurl.com}, with over 42\% (or 1.18million) of the total URLs. 
We investigated the tinyurl links, and found that 99.9\% of them  pointed  to only two  destinations: one was a webpage that had been blocked by tinyurl, which means tinyurl had classified it as malicious or spam; the other is a known phishing webpage \url{www.facebook-goodies.com}\footnote{This website has now been deleted, but it is available through Wayback Machine, arguably the most comprehensive digital archive	of the World Wide Web.}. 
By performing a content analysis, we found that the
vast majority of all the URLs could be clustered into only two distinct spam campaigns.

It is almost certain that the Bursty botnet was carefully designed and centrally controlled for the purpose of a spamming attack. A number of tricks have been used to hide the attack. 

Firstly, the bots were created in large numbers, and each bot was used    to generate   a small number of tweets in the first few minutes only and then the bots all became silent. Most existing bot detection methods are not able to identify such %
inactive bots. 

Secondly, the bots used a complex network of URL shorteners and
redirects to obfuscate the final landing pages, such that the vast number of URLs were used only once, which could effectively evade most spam filters. Also it was not easy for users to tell on the final destination of the URLs. 

Thirdly,  the botnet  directly targeted over 1.3m distinct Twitter users by mentioning their usernames, which significantly increased the chance of the URLs being clicked.  Our analysis revealed just how successful this technique was: on average over 61\% of the posted URLs that lead to a phishing campaign
were clicked, which could yield  a remarkable revenue   by selling stolen personal
data. 

With the above information and further research, we were even able to  track down the
alleged botmaster of the Bursty Botnet.
Our detailed analysis on the spamming attack  of the Bursty botnet will be published in another paper.

\section{Reflection on Twitter bot detection }

\subsection{Failure of existing detection methods}
In recent years there have been many efforts to detect Twitter bots. Some have produced plausible results. Most of them  relied on  heuristic assumptions on `common' features that should supposedly be shared by all bots.      
It is clear that these assumed features are not shared by all bots. For example the Bursty botnet and the Star Wars botnet show some properties that are diametrically different from those assumptions. As a result the the existing methods have not been successful in detecting large botnets. We have verified that the Bursty bots and the Star Wars bots can fool one of the latest and more advanced bot detection tools \cite{davis_botornot:_2016}.  

One reason was that previous studies were restricted by the lack of ground truth data. Since the available datasets all contained mixed types of bots, researchers had to search for and focus on general features shared by all bots in the datasets. 

Another possible reason was that the previously studied features of bots have been 'fading', because most of these features  could be easily avoided in the design of later botnets as botmasters  must have closely followed the  development on bot detection. 
%
\subsection{A long-term battle with no silver bullet}

The Bursty botnet and the Star Wars botnet exhibit distinctive properties that have been overlooked so far,   namely  the tweet location distribution and the bursty tweeting behaviour. 
But we do not expect that further study on these features will lead to discovery of many other new botnets. There is a strong incentive for botmasters to deliberately create new botnets that do not show any of the features that have already been `exposed' by researchers. Ideally   botmasters would  create new botnets that are completely different from existing ones.  

Indeed, we expect the battle on bot detection to be a long-term process, where researchers have to keep proposing new detection methods  to catch up with new generations of botnets, which are likely to become ever more deceptive. 
As such, although it has been highly desired for by the research community, we do not believe it will be possible  to  develop a `generalised'  method  to detect all types of bots.

\section{Conclusion}
 		 
The discovery of the Bursty botnet and the Star Wars botnet  provided valuable ground truth data for the Twitter bot research community. 
Both botnets are unusually large. Each contains hundreds of thousands of bots. They are different from other datasets because each of the botnets contains a single network of bots that exhibit the same properties and were created and controlled by the same botmaster. It seems Twitter has removed the Star Wars botnet since our publication~\cite{echeverria_star_2017}. As of this writing, most of the 500,000 Bursty bots are still alive on   Twitter. Researchers can collect   them by following instructions in this paper. Researchers, however, should hurry to collect them before Twitter deletes these accounts too.

It is interesting to point out that the Bursty Botnet and the Star Wars Botnet were discovered by their unusual tweeting behaviours, which is in a rather `unconventional' way that was different from previous bot detection efforts. 
 This also means it does not suffer the biases
induced by relying on Twitter's black-box suspension algorithm or URL blacklisting
services, which makes it a valuable ground truth for future research.

These new datasets not only enabled us to reflect on the previous assumptions and detection methods, but also provided us a rare and valuable opportunity to investigate how Twitter bots were designed, created, and used for a spamming attack in the cyberspace.

\bibliographystyle{IEEEtran}



\end{document}